# Experimental Evaluation of Post-Quantum Homomorphic Encryption for Privacy-Preserving I2I Communication in ITS

Abdullah Al Mamun, *Student Member, IEEE*, Kyle Yates, Antsa Pierrottet, Mashrur Chowdhury, *Senior Member, IEEE*, Ryann Cartor, Shuhong Gao

***Abstract*— This study experimentally evaluates the feasibility of post-quantum secure Homomorphic Encryption (HE) for privacy-preserving Infrastructure-to-Infrastructure (I2I) communication in Intelligent Transportation Systems (ITS). Unlike prior simulation-based efforts, this work implements three lattice-based HE schemes: Brakerski-Fan-Vercauteren (BFV), Brakerski-Gentry-Vaikuntanathan (BGV), and Cheon-Kim-Kim-Song (CKKS), within a real experimental pipeline representing roadside unit (RSU)-Cloud data exchange over Wi-Fi and Ethernet networks. The experiments benchmark encrypted addition and addition-plus-multiplication operations representing key analytical tasks, such as vehicle queue assessment and regional speed computation. Results show that while BFV achieves sub-5-second latency suitable for intersection-level analytics, BGV supports regional aggregation with 10 to 30-second updates. CKKS, though exhibiting higher latency (21-32 seconds), remains practical for minute-scale applications like eco-driving. These findings demonstrate that post-quantum HE can enable privacy-preserving ITS backhaul analytics when latency requirements align with application needs. The study also presents optimization pathways, including algorithmic tuning, network adaptation, and hardware acceleration, to reduce end-to-end delay.**

***Index Terms*— Homomorphic encryption, Intelligent transportation systems, Network latency, Post-quantum cryptography, Privacy protection, Roadside units.**

## I. INTRODUCTION

INTELLIGENT Transportation Systems (ITS) integrate advanced information and communication technologies into transportation infrastructure and vehicles to improve traffic efficiency, enhance road safety, and reduce congestion. These systems rely on continuous data exchange between vehicles, roadside units (RSUs), transportation infrastructure, and cloud services, enabling applications, such as collision avoidance, queue monitoring, speed advisory systems, and route optimization [1]. This data exchange is primarily established through Vehicle-to-Everything (V2X) communication [1], [2]. V2X includes two key modes of interaction: Vehicle-to-Vehicle (V2V) communication, which supports direct message exchange between vehicles to enhance safety and coordination, and Vehicle-to-Infrastructure (V2I) communication, which enables vehicles to interact with RSUs and traffic control systems. In parallel, Infrastructure-to-Infrastructure (I2I) communication serves as a complementary layer that connects RSUs, traffic management centers, and cloud platforms for large-scale data aggregation and analytics. Although I2I is not formally part of the V2X framework, it plays a critical role in supporting V2X-enabled ITS applications through backend coordination and distributed data processing. Together, these communication layers enable seamless and cooperative operations within modern ITS environments.

Despite its benefits, the broadcast nature of V2X communication introduces serious privacy and security concerns. Basic Safety Messages (BSMs) are transmitted in plaintext as defined by the Society of Automotive Engineers (SAE) J2735 standard [3]. Each message includes a set of attributes, such as timestamp, geographic position, speed, and heading. To ensure authenticity and integrity, the Security Credential Management System (SCMS) digitally signs every BSM, but the messages themselves are not encrypted [4]. RSUs receive these BSMs, verify their signatures, and then forward the verified data to centralized or cloud-based systems that perform higher-level analytics, such as traffic-flow harmonization, congestion monitoring, travel-time estimation, or incident detection [5], [6]. When transmitted in plaintext, these data streams expose vehicle trajectories and driving patterns that can be reconstructed through temporal and spatial correlation. Although pseudonymous certificates reduce the possibility of direct identity disclosure, adversaries or untrusted cloud servers can still link consecutive messages from the same vehicle using location and speed attributes [7], [8].

Even when RSUs apply conventional or post-quantum encryption before forwarding data to cloud or data analytics centers (data centers), these systems must decrypt the information before performing analytics. This re-exposes sensitive vehicular details during computation and creates potential privacy vulnerabilities, particularly when such infrastructure is managed by third-party or external entities. These challenges highlight a critical gap in the current ITS

This work was supported in full by the National Center for Transportation Cybersecurity and Resiliency (TraCR) (a U.S. Department of Transportation National University Transportation Center) under Grant 69A3552344812 and Grant 69A3552348317.

Abdullah Al Mamun and Mashrur Chowdhury are with the Glenn Department of Civil Engineering, Clemson University, Clemson, SC 29634 USA (e-mail: abdullm@clemson.edu; mac@clemson.edu).

Kyle Yates, Antsa Pierrottet, Ryann Cartor and Shuhong Gao are with the School of Mathematical and Statistical Sciences, Clemson University, Clemson, SC 29634 USA (e-mail: kjyates@clemson.edu; arakoto@clemson.edu, rcartor@clemson.edu, and sgao@clemson.edu).

security framework. Traditional encryption techniques, such as the Advanced Encryption Standard (AES) [9], can protect data at rest and during transmission but cannot maintain confidentiality while the data is being processed. Similarly, post-quantum cryptographic algorithms standardized by the National Institute of Standards and Technology (NIST) [10], still require decryption before analysis. As a result, vehicle-specific trajectories and behavioral information remain exposed to the computational entity during analytics, reinforcing the need for privacy-preserving solutions within the RSU-Cloud or RSU-Data Center communication path.

Homomorphic Encryption (HE) provides a potential solution to this limitation because it enables computation directly on encrypted data without decryption [11]. Within an ITS framework, RSUs can encrypt aggregated BSM data with HE schemes before transmitting it to the cloud or data analytics center. The cloud/data analytics center can then perform data analytics on encrypted data without access to the underlying plaintext. The RSU, which retains the private key, decrypts only the final results. This mechanism allows privacy-preserving analytics while maintaining compatibility with existing SCMS-based authentication that secures V2V and V2I communication.

While HE enables privacy-preserving computation, its use in this study is intentionally limited to the I2I communication link rather than the V2V or V2I broadcast links. Privacy-critical V2X messages, particularly BSMs, must be transmitted in plaintext at a fixed rate of 10 Hz (10 BSMs per second) as required by SAE J2735 and J2945/1 to ensure immediate situational awareness among nearby vehicles [3], [12]. Encrypting these messages, especially using post-quantum HE schemes due to their large ciphertext sizes, would violate this mandatory broadcast requirement by adding latency and causing packet fragmentation. For this reason, HE cannot be used within the V2X broadcast channel without disrupting the operations of vehicle-level ITS safety applications. In contrast, I2I communication for non-real-time ITS applications can operate on aggregated data streams and does not impose strict real-time constraints, making it a technically appropriate and standards-compliant communication link for evaluating the feasibility of post-quantum HE–based analytics for many ITS applications.

This study investigates the practical feasibility of applying post-quantum HE to I2I communication, specifically within the RSU-Cloud or RSU-Data Center data exchange pipeline. The experimental evaluation focuses on representative HE operations that commonly underline data aggregation and analytical tasks in ITS applications. These operations include encrypted addition and encrypted multiplication, which correspond to fundamental computations, such as vehicle counting, congestion estimation, and average speed calculation. Four operation configurations were tested to evaluate scalability and network impact: 49 additions (with 50 encrypted inputs), 99 additions (100 encrypted inputs), 199 additions (200 encrypted inputs), and 199 additions combined with one multiplication (200 encrypted inputs). Each configuration was executed under two network environments, Wi-Fi and Ethernet, to capture realistic transmission characteristics, latency behavior, and computational performance under varying data volumes.

Unlike previous ITS-focused HE studies that rely primarily on simulations or theoretical models [13], [14], [15], [16], [17], this work conducts end-to-end experimentation using actual hardware and real network links, enabling an assessment of how HE impacts computational latency, communication overhead, ciphertext size, and throughput under realistic network conditions. To this end, we benchmark three leading lattice-based HE schemes, i.e., Brakerski-Fan-Vercauteren (BFV) [18], Brakerski-Gentry-Vaikuntanathan (BGV) [19], [20], and Cheon-Kim-Kim-Song (CKKS) [21], across both wireless and wired I2I communication channels. These schemes were selected because they are widely adopted, support both exact (BFV, BGV) and approximate (CKKS) arithmetic, and are based on the Ring Learning With Errors (RLWE) problem [22], which is believed to be secure against quantum adversaries. In contrast, earlier ITS applications commonly employed partially homomorphic encryption (PHE) schemes such as Paillier [23], [24], [25], [26], [27], [28], [29], which support only homomorphic addition between ciphertexts and scalar multiplication by a constant, but lack homomorphic multiplication between encrypted messages. Furthermore, their security relies on the integer factorization problem, making them vulnerable to quantum attacks [30]. Our work thus presents the first simulation-free benchmarking of post-quantum secure HE schemes for ITS, combining real-world network transmission with representative ITS computations to evaluate practical feasibility in latency-sensitive vehicular environments.

## II. BACKGROUND AND LITERATURE REVIEW

This section presents the foundations for the HE schemes used in our experiments and reviews prior work that applied HE to ITS domains. It also outlines the mathematical structure and parameter choices essential for understanding the implementation and security guarantees of the selected schemes.

### *A. Overview of Homomorphic Encryption Schemes*

A concise, high-level overview of the BGV, BFV, and CKKS homomorphic encryption schemes is provided to establish the foundational concepts for this study. BGV and BFV support encrypted computation for exact arithmetic (e.g., finite field arithmetic), while CKKS supports arithmetic for floating-point numbers to some precision accuracy. A HE scheme consists of a collection of the following algorithms:

- $KeyGen$ generates the secret key $sk$, the public key $pk$, and an evaluation key $ek$.
- $Encrypt$ encrypts a message $\boldsymbol{m}$ as a ciphertext $ct$ using the public key $pk$.
- $Decrypt$ decrypts a ciphertext $ct$ to a message $\boldsymbol{m}$ using the secret key $sk$.

- $Add$ performs an addition between two ciphertexts $ct_1$ and $ct_2$.
- $Multiply$ performs a multiplication between two ciphertexts $ct_1$and $ct_2$ using the evaluation key $ek$.
- $Modswitch$ performs a scaling on a ciphertext to control encryption noise or bit expansion. This is typically performed immediately after a multiplication.
- In addition to the above algorithms, CKKS also includes $Encode$ and $Decode$ steps, which maps vectors of floating-point numbers to integer polynomials using a scaling factor $\Delta$ to preserve some precision accuracy.

The attractive feature of these HE schemes is that performing $Add$ or $Multiply$ on ciphertexts results in the same value as performing the same operations on the original messages and then encrypting. For instance, if $ct_1$ is an encryption of message $\boldsymbol{m_1}$ and $ct_2$ is an encryption of message $\boldsymbol{m_2}$, homomorphically adding $ct_1$ and $ct_2$ results in an encryption of $\boldsymbol{m_1} + \boldsymbol{m_2}$ (or $\boldsymbol{m_1} \times \boldsymbol{m_2}$ for $Multiply$). Furthermore, these operations can be performed by anyone with access to the public key and ciphertexts without needing or leaking any private information. These operations directly on ciphertexts are called homomorphic operations.

We omit the finer mathematical details of the described schemes and refer the reader to [18], [19], [20], [21], [31], [32], [33] for a more comprehensive view of HE foundations. However, we briefly present the mathematical structure of ciphertexts used in BGV, BFV, and CKKS. Encrypted messages in these schemes take the form of polynomial pairs $(\boldsymbol{a}, \boldsymbol{b})$ over the ring**:**

$$R_q = \mathbb{Z}_q[x]/(x^n + 1)$$

where $n$ is a power of two and $q$ is the ciphertext modulus. Let $\boldsymbol{s} \in R_q$ be a small secret key and $\boldsymbol{a}$ be a uniformly random polynomial in $R_q$. The ciphertext component $\boldsymbol{b}$ is computed differently depending on the scheme:

- BGV: $\boldsymbol{b} = -\boldsymbol{as} + \boldsymbol{m} + t\boldsymbol{e} \quad mod(x^n + 1, q)$
- BFV: $\boldsymbol{b} = -\boldsymbol{as} + \lfloor q/t \rceil \boldsymbol{m} + \boldsymbol{e} \quad mod(x^n + 1, q)$
- CKKS: $\boldsymbol{b} = -\boldsymbol{as} + \boldsymbol{m} + \boldsymbol{e} \quad mod(x^n + 1, q)$

Here, $\boldsymbol{m}$ is the plaintext message (or its encoded polynomial), $\boldsymbol{e}$ is a small noise polynomial with integer coefficients, and $t$ is the plaintext modulus used in BFV and BGV. For CKKS, the plaintext $\boldsymbol{m}$ is typically a polynomial obtained from encoding a vector of real-valued messages, and the scheme tolerates approximate arithmetic. Choosing the parameters $n, q$, and $t$ is important for determining the appropriate security level and ciphertext size, which are discussed in the following subsection.

### *B. Parameter Selection and Security*

We use the OpenFHE implementation of BFV, BGV, and CKKS [34], following the 2024 security guidelines proposed in [35]. Parameters are chosen to ensure 128-bit post-quantum security (NIST Security Level 1) and efficient real-time operation on ITS edge hardware. Table I summarizes the configuration of each scheme used in our experiments. The "Levels" column indicates the multiplicative depth of the homomorphic arithmetic circuit: a level of 1 permits only homomorphic additions (no ciphertext-ciphertext multiplications), while a level of 2 supports one such multiplication.

TABLE I
HE PARAMETERS (128-BIT SECURE) USED IN THIS STUDY
(SELECTED BASED ON [35])

| Scheme | $n$ | $\log_2(q)$ | $t$ | $\log_2(\Delta)$ | Levels | Public Key Size (bytes) | Ciphertext Size (bytes) |
|---|---|---|---|---|---|---|---|
| BFV | 4096 | 106 | 65537 | N/A | 1 | 131895 | 131939 |
| BGV | 8192 | 106 | 65537 | N/A | 1 | 656789 | 394573 |
| CKKS | 16384 | 106 | N/A | 38 | 1 | 1312151 | 787791 |
| CKKS | 16384 | 106 | N/A | 38 | 2 | 1574473 | 1050129 |

Note- N/A: Not Applicable; Levels = Multiplicative Depth + 1.

### *C. Current Use of Homomorphic Encryption in ITS*

As introduced earlier, previous ITS-focused HE studies primarily used PHE such as Paillier and relied on simulations or theoretical models. While these efforts demonstrated the conceptual feasibility of privacy-preserving computations, they lacked real-world validation and relied on cryptosystems vulnerable to quantum attacks.

In contrast, somewhat homomorphic encryption (SHE) and fully homomorphic encryption (FHE) schemes, such as BGV, BFV, and CKKS, offer broader functionality. SHE or leveled schemes support a limited number of additions and/or multiplications on ciphertexts, while FHE supports an unlimited number of operations through bootstrapping [18], [36]. These schemes are based on hard lattice problems, making them resistant to quantum attacks. All practical SHE/FHE schemes to date are considered post-quantum secure [36]. A growing body of research [14], [15], [16], [17], [37], [38], [39], [40], [41], [42] has investigated SHE and FHE for ITS, including implementations using HE libraries like SEAL [37], [38], [39], Pyfhel [14], [40], Helib [38], and PALISADE [16]. Other studies leverage the TFHE scheme [43] for ITS [17], propose novel constructions [15], or survey HE-based methods [42]. Related approaches, such as secure multi-party computation and federated learning, have also been explored for ITS applications [44], [45], [46].

As discussed in the Introduction section, our study is the first to experimentally evaluate SHE for ITS using real communication links, rather than through simulation or theoretical design. It is also the first study to conduct such experiments using the OpenFHE library.

## III. EXPERIMENTAL SETUP

This section presents the experimental framework for benchmarking post-quantum secure HE in an I2I pipeline. The focus is on the feasibility of encrypted computation for RSU-backhaul analytics that protects vehicle information during processing.

### *A. Concept and Rationale*

Fig. 1 illustrates the proposed privacy-preserving framework that integrates HE within an I2I communication pipeline. As mentioned in the Introduction section, in a conventional ITS architecture, RSUs collect BSMs broadcast in plaintext by connected vehicles, verify their authenticity, and forward the verified data to centralized computational entities, such as cloud or data analytics centers for higher-level computations. However, plaintext or decrypted transmissions can expose sensitive vehicular information, such as location, trajectory, or behavioral patterns, to untrusted computational entities. Even when encryption is applied for transport security, data must eventually be decrypted for analysis, reintroducing privacy risk.

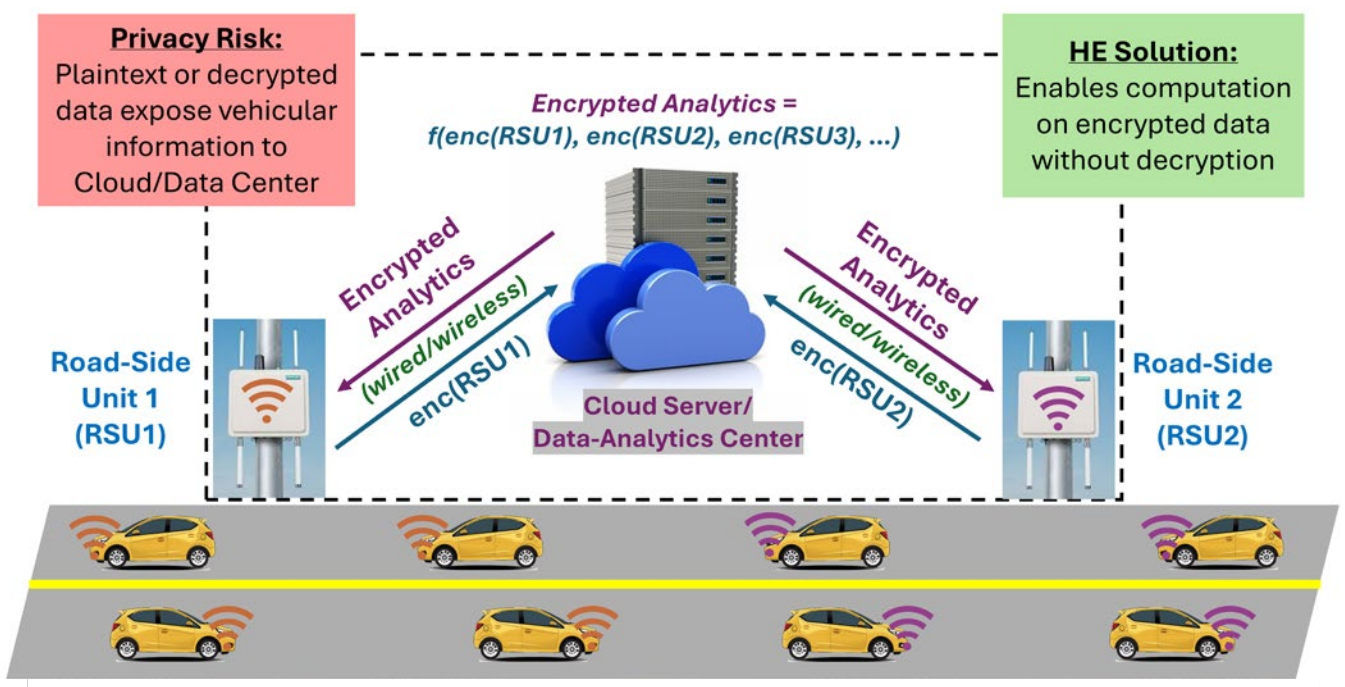

Fig. 1. Privacy-preserving I2I data-exchange framework using HE between RSUs and a cloud or data analytics center. Abbreviations- HE: homomorphic encryption; I2I: infrastructure-to-infrastructure; enc(RSU1): encrypted data from RSU1.

Homomorphic Encryption provides a privacy-preserving alternative by enabling computations directly on ciphertexts without requiring decryption. In the illustrated workflow (Fig. 1), each RSU encrypts local BSM aggregates before forwarding them to the cloud or data analytics center. The cloud/data analytics center performs operations on encrypted data and returns only ciphertext results. Decryption is performed solely by the RSU, which holds the private key, ensuring that the computational platform remains blind to the plaintext data. This architecture enables privacy-preserving analytics while maintaining interoperability with existing ITS data exchange standards and communication protocols.

### *B. Operation-Centered Benchmark Design*

The experimental design focuses on fundamental encrypted computations that occur across multiple ITS analytics tasks. Instead of binding the evaluation to a specific application, the study generalizes the benchmarking process around core HE operations that characterize typical RSU-Cloud or RSU-Data Center data exchanges. Four operation configurations were tested to assess scalability and latency: (1) 49 encrypted additions with 50 encrypted inputs, (2) 99 encrypted additions with 100 encrypted inputs, (3) 199 encrypted additions with 200 encrypted inputs, and (4) 199 encrypted additions with one encrypted multiplication (200 encrypted inputs). These configurations represent the essential building blocks for operations, such as counting, summation, and averaging.

### *C. RSU-Cloud Encrypted Analytics and ITS Application Mapping*

The RSU-Cloud (or RSU-Data Center) encrypted-analytics framework supports a variety of ITS applications that require secure processing of sensitive information. The supported applications are mapped to established service packages defined in the United States Department of Transportation (USDOT) Architecture Reference for Cooperative and Intelligent Transportation (ARC-IT) framework [47]. Table II lists representative ARC-IT service packages whose analytical workflows can be securely supported by the proposed RSU-Cloud encrypted-analytics framework. Each application involves linear statistical computations, i.e., counts, sums, or averages, that map efficiently to additive and low-depth multiplicative homomorphic operations.

The operations summarized in Table II represent the majority of computational tasks currently performed in transportation management and data analytics centers, whether hosted in trusted public-sector facilities or untrusted commercial cloud environments. In both settings, most functions rely on linear statistics, such as counts, summations, and averages. Additive homomorphism enables secure encrypted counting, while limited multiplicative depth supports computation of averages and rate-based metrics. Accordingly, the BFV and BGV schemes are well suited for exact integer-based measurements, whereas the CKKS scheme is more appropriate for continuous traffic variables where small approximation errors are acceptable.

It is to be noted that certain applications that require comparisons or threshold detections, such as identifying minimum travel times, determining queue thresholds, or verifying compliance cut-offs, necessitate deeper homomorphic circuits (i.e., FHE or fast-bootstrapping schemes) or specialized secure-comparison protocols. These operations extend beyond the computational depth evaluated in this study but represent promising directions for future enhancements of the RSU-Cloud encrypted-analytics framework.

TABLE II
REPRESENTATIVE ARC-IT SERVICE PACKAGES AND CORRESPONDING HE OPERATIONS SUPPORTED BY THE RSU-CLOUD ENCRYPTED-ANALYTICS

| ARC-IT Service Package | ITS Application | Primary Analytical Task | Required HE Operations |
|---|---|---|---|
| TM03: Traffic Signal Control [48] | Intersection and corridor signal optimization | Aggregation of RSU-level presence indicators and delay metrics | Encrypted summation of *enc*(0)/*enc*(1) presence indicators |
| TM17: Speed Warning and Enforcement [49] | Speed monitoring and compliance evaluation | Segment-level speed profiling and compliance-rate computation | Encrypted summation and averaging (sum + scalar multiplication) |
| TM21: Speed Harmonization [50] | Dynamic speed advisory generation | Weighted or moving averages of aggregated vehicle speeds | Encrypted summation and scalar multiplication |
| TM02: Vehicle-Based Traffic Surveillance [51] | Travel-time and flow estimation using probe data | Aggregation of link travel-time samples and flow counts | Encrypted summation and averaging |
| MC06: Work Zone Management [52] | Queue detection and speed-compliance monitoring | Encrypted presence counts and mean-speed computation | Encrypted summation and averaging |
| TM07: Regional Traffic Management [53] | Multi-RSU and corridor-level coordination | Cross-RSU aggregation and rolling-average generation | Encrypted summation and averaging |

### *D. Experimental Workflow and Processing Steps*

To evaluate the applicability of HE in real-world ITS environments, the experimental framework emulates I2I communication between a RSU and a Cloud or Data Analytics Center. In this configuration, the RSU (sender) encrypts and transmits aggregated vehicular data to the Cloud or Data Center (receiver), which performs computations directly on the ciphertexts without decryption. This setup replicates a practical RSU-Cloud data pipeline, where the Cloud may be operated by either a trusted public-sector entity or an untrusted commercial provider.

Two representative computation scenarios were implemented to assess the computational and communication feasibility of HE within this pipeline. The first scenario evaluates addition-only operations that emulate ITS applications, such as congestion counting and encrypted speed aggregation. All three lattice-based schemes, i.e., BFV, BGV, and CKKS, were implemented using the OpenFHE C++ library to benchmark their performance under lightweight additive workloads. In this configuration, the RSU decrypts the returned ciphertext sum and may perform simple plaintext post-processing, such as computing averages or applying threshold decisions. This reflects realistic ITS edge analytics, where local processing at trusted RSUs complements encrypted computation at the cloud layer while ensuring privacy preservation. The second scenario includes both addition and multiplication, executed using only the CKKS scheme, which supports approximate fixed-point arithmetic. In this case, averaging is carried out entirely within the encrypted domain by the Cloud/Data Center, eliminating the need for intermediate decryption. The BFV and BGV schemes were excluded from this configuration because they lack native support for efficient ciphertext rescaling and precision-preserving fixed-point multiplication, which are two key requirements for accurate arithmetic in encrypted averaging tasks.

Both scenarios were evaluated under two network conditions, i.e., Wi-Fi and Ethernet, to measure the impact of wireless and wired infrastructures on computational delay, communication overhead, and reliability. All implementations used OpenFHE's modular APIs for key generation, parameter tuning, encryption, serialization (fragmentation), and ciphertext transmission. The full workflow, summarized in Table III, describes the sequence of sender and receiver operations, from key generation to metric logging.

### *E. Experimental Parameters and System Configuration*

Beyond the cryptographic parameters described in Table I, additional system-level and runtime configurations were applied to ensure stable, low-latency operation under both wired and wireless backhaul conditions. These configurations were selected to emulate realistic ITS environments in which RSUs exchange encrypted data with centralized cloud or data-center servers. Table IV summarizes the complete system-level parameters used in the experimental setup.

TABLE III
EXPERIMENTAL WORKFLOW

| Step | Sender (RSU) | Receiver (Cloud/Data Center) |
|---|---|---|
| 1. Context and Key Generation | Generate the OpenFHE context and a public/private key pair using selected parameters for BFV, BGV, or CKKS. Share the public key with the receiver while securely storing the secret key. | Receive the context and public key to enable encryption and homomorphic operations. |
| 2. Message Preparation and Encryption | Prepare plaintext messages (e.g., binary 0/1 or speed values). Encrypt data using the public key and encode via OpenFHE APIs. | Encrypt simulated data for multiple vehicles (in RSUs) using the same public key. |
| 3. Serialization | Serialize encrypted ciphertexts into binary format using OpenFHE's serialization tools. | - |
| 4. Fragmentation | Fragment serialized ciphertexts into ≤1400 byte UDP packets with prepended sequence numbers, ensuring compatibility with Wi-Fi/Ethernet MTUs. | - |
| 5. Transmission over Network | Transmit encrypted UDP packet fragments over the selected medium (Wi-Fi for wireless or Ethernet for wired). | Receive and buffer incoming UDP fragments for reassembly. |
| 6. Reassembly and Deserialization | - | Reassemble fragments using sequence numbers and deserialize to reconstruct the original ciphertext. |
| 7. Homomorphic Computation | - | Perform encrypted aggregation on received ciphertexts. For the addition-only scenario, apply homomorphic summation. For the average-speed scenario, use the CKKS scheme to execute both addition and approximate scalar multiplication to derive encrypted averages. |
| 8. Serialization and Fragmentation | - | Serialize and fragment the homomorphic computation results (ciphertext). |
| 9. Return Transmission | Receive result (ciphertext) fragments. | Transmit encrypted result fragments (ciphertext) back to the sender over Wi-Fi or Ethernet. |
| 10. Decryption and Output | Reassemble and deserialize received ciphertext and decrypt using private key. | - |
| 11. Final Output and Metrics | Perform any final plaintext-side computations if needed and log performance metrics. | Log performance metrics. |

A packet size of 1400 bytes was selected to remain below the standard Maximum Transmission Unit (MTU) limits for Ethernet and Wi-Fi (typically 1500 bytes). This constraint minimizes IP-layer fragmentation and packet loss during reassembly. The choice of Wi-Fi and Ethernet as communication media reflects the connectivity options supported by commercial RSUs. For example, the commercial Cohda Wireless MK6 RSU supports multiple network interfaces, including Gigabit Ethernet (IEEE 802.3), Wi-Fi (IEEE 802.11 a/b/g/n/ac), and cellular LTE/5G for wide-area backhaul connectivity [54].

Wi-Fi-based UDP transmission was employed to emulate the wireless backhaul commonly used in RSU infrastructure and to assess jitter, packet pacing, and reassembly reliability under time-sensitive conditions. To analyze scalability, three data-volume levels were tested: 50, 100, and 200 encrypted inputs. In the addition-only scenario, encrypted binary presence indicators (*enc*(0)/*enc*(1)) were used, while the addition-plus-multiplication configuration employed encrypted speed values randomly selected between 40 and 60 mph to evaluate the feasibility of CKKS-based homomorphic averaging.

Transmission pacing was empirically tuned to approximately 1.2 × RTT for Wi-Fi, where RTT stands for round-trip time, and 100 ms for Ethernet to prevent packet drops caused by socket buffer saturation. A 5-second receive timeout between consecutive fragments was configured to prevent indefinite blocking during data reassembly and to emulate the responsiveness expected in real-time ITS backhaul communications. This ensures that if a UDP fragment is delayed or lost, the receiver proceeds without halting the entire session. All computations were executed sequentially on single threads to replicate the processing limitations usual in RSUs and other edge devices. Each test was repeated multiple times to ensure statistical consistency: 20 trials for BFV and BGV, and 5 trials for CKKS.

### *F. Performance Metrics*

Performance was evaluated using several key metrics relevant to I2I communication, including end-to-end computational latency, communication latency (RTT – computational latency), jitter, number of UDP fragments, and times for packet fragmentation, packet reassembly, encryption, and decryption. Accuracy was also assessed by comparing decrypted outputs of homomorphic computations with their plaintext equivalents to confirm the correctness of encrypted processing. Together, these measurements provide a comprehensive evaluation of the feasibility of deploying HE schemes in real-time, near-real-time, and offline ITS applications.

## IV. RESULTS AND DISCUSSION

The experimental analysis highlights a key challenge for privacy-preserving I2I systems: communication latency dominates performance bottlenecks across all tested HE schemes. Although the chosen parameters achieve 128-bit post-quantum security with relatively compact ciphertexts (Table I), the substantial data expansion from encryption introduces significant transmission/communication overhead, outweighing computational differences among schemes. The decrypted results from all experiments were verified against their plaintext equivalents, confirming 100% accuracy of the homomorphic computations. To systematically examine these constraints and trade-offs, the results are presented in three parts: (i) computation latency analysis (Table V), (ii) network-level communication characteristics (Table VI), and (iii) comparative visual analysis (Fig. 2 and Fig. 3) showing latency and homomorphic operation times across HE schemes.

### *A. Computation Latency*

Table V summarizes the total end-to-end computation latency of the BFV, BGV, and CKKS schemes under Wi-Fi and Ethernet configurations. Each entry represents a specific combination of data volume, HE scheme, and network medium, with measured times for encryption, homomorphic computation, decryption, and overall processing.

TABLE IV
EXPERIMENTAL PARAMETERS AND SYSTEM CONFIGURATION

| Category | Parameter | Value/Setting |
|---|---|---|
| Fragmentation | Maximum Transmission Unit (MTU) | 1400 bytes |
| Transmission delay (delay between successive packet transmissions) | Wi-Fi delay | Approximately 1.2*RTT without any packet fragment loss (upto ~4,500 ms for BFV; ~13,000 ms for BGV; and ~30,000 ms for CKKS) |
| | Ethernet delay | 100 ms |
| Networking | Socket buffer size | 4 MB (sender and receiver) |
| | Receive timeout (between fragments) | 5 s |
| Execution | Threading model | Single-threaded (sequential) |
| Number of trials for each scenario | BFV, BGV, and CKKS | 20, 20, and 5 |
| Sender Device | Processor | 13th Gen Intel Core i9-13900HX @ 2.2 GHz (base) |
| | CPU cores | 24 cores (8 performance + 16 efficiency) |
| | RAM | 32 GB |
| | Wi-Fi adapter / Protocol | Intel Wi-Fi 6 AX201 160MHz / 802.11n |
| | Ethernet adapter / Protocol | Killer E3100G 2.5GbE / IEEE 802.3ab |
| Receiver Device | Processor | Intel Core i7-5500U @ 2.40GHz (base) |
| | CPU cores | 2 cores / 4 threads |
| | RAM | 8 GB |
| | Wi-Fi adapter / Protocol | Intel Dual Band Wireless-AC 7265 / 802.11n |
| | Ethernet adapter/ Protocol | Cisco AnyConnect Virtual Adapter / IEEE 802.3ab |
| Network Environment | Wi-Fi transmission/receiving speed | 300 Mbps |
| | Ethernet transmission/receiving speed | 1 Gbps |

The results indicate that while encryption and decryption times remain modest across all schemes (ranging from approximately 5 to 25 ms), the dominant cost lies in the homomorphic operation times, which scale nearly linearly with the number of encrypted inputs. For instance, in the BFV scheme (Wi-Fi), increasing the data volume from 50 to 200 encrypted inputs results in a ~3.5 times increase in computation latency (189.72 ms to 680.98 ms).

The CKKS scheme exhibits the highest computational overhead in all configurations. Even without multiplicative operations, CKKS latency rises from ~685 ms at 50 encrypted inputs to over 2800 ms at 200 encrypted inputs under Wi-Fi. When a single multiplication is included (to enable homomorphic averaging), the latency increases further, reaching 3210.93 ms over Wi-Fi and 3069.82 ms over Ethernet for 200 encrypted inputs. This increase reflects CKKS's rescaling and precision-handling requirements, which introduce additional processing cost. The larger polynomial ring size used in CKKS (n=16384), compared with BGV (n=8192) and BFV (n=4096), also contributes to the higher computational overhead despite providing greater numerical precision and security. Overall, both Wi-Fi and Ethernet configurations yield similar computational latency trends, confirming that these results primarily reflect computation time independent of network transmission effects, which are analyzed next.

*B. Communication Latency and Jitter*

Table VI presents communication-level latency, RTT, jitter, packet fragmentation, and packet reassembly characteristics for all HE configurations. These results reveal how ciphertext expansion impacts data transmission performance in RSU-Cloud or RSU-Data Center communication over different network media.

Baseline plaintext transmission, with 4-6 byte messages, achieves RTTs of 3.8-4.1 ms under both Wi-Fi and Ethernet with zero fragmentation overhead. In contrast, HE ciphertexts expand dramatically. For example, a BFV ciphertext of ~132 KB generates 95 fragments for 50 encrypted inputs, while CKKS ciphertexts for addition-plus-multiplication workloads (~1 MB) require over 750 fragments. This fragmentation leads to RTTs exceeding 24,000 ms (24 s) and total communication latencies surpassing 21,700 ms (21.7 s) in some configurations, underscoring the critical impact of ciphertext size on transmission efficiency in I2I systems.

Notably, despite the fundamental differences between Wi-Fi and Ethernet, such as Wi-Fi's shared, interference-prone wireless spectrum versus Ethernet's dedicated, full-duplex wired channel, and their respective disparities in bandwidth, jitter, and error rate, the total communication latency across schemes remains nearly identical in both mediums. This counterintuitive outcome can be attributed to the fact that fragmentation, serialization, UDP socket buffering, and inter-packet pacing dominate the delay pipeline, thereby neutralizing the practical benefits of Ethernet's higher bandwidth.

TABLE V

END-TO-END COMPUTATIONAL LATENCY OF HE SCHEMES ACROSS VARYING ENCRYPTED INPUTS AND COMMUNICATION MEDIA

| Scheme | Encrypted Inputs (Number of HE operations) | Medium | Encryption time at sender (ms) | Time to perform HE operations at receiver (ms) | Decryption time at sender (ms) | Mean computational end-to-end latency (ms) |
|---|---|---|---|---|---|---|
| BFV (addition only) | 50 (49 additions) | Wi-Fi | 6.40 ± 2.04 | 189.72 ± 14.87 | 4.60 ± 0.88 | 200.72 |
| | | Ethernet | 5.84 ± 2.12 | 183.56 ± 4.19 | 4.36 ± 0.85 | 193.76 |
| | 100 (99 additions) | Wi-Fi | 5.44 ± 2.14 | 347.61 ± 6.80 | 4.73 ± 1.08 | 357.78 |
| | | Ethernet | 5.99 ± 2.14 | 343.00 ± 7.57 | 4.78 ± 0.96 | 353.77 |
| | 200 (199 additions) | Wi-Fi | 6.54 ± 1.96 | 680.98 ± 37.33 | 4.94 ± 0.87 | 692.46 |
| | | Ethernet | 6.40 ± 1.34 | 675.37 ± 5.95 | 4.97 ± 0.63 | 686.73 |
| BGV (addition only) | 50 (49 additions) | Wi-Fi | 10.19 ± 2.24 | 346.93 ± 12.24 | 8.16 ± 2.29 | 365.27 |
| | | Ethernet | 9.70 ± 2.07 | 344.17 ± 14.31 | 7.36 ± 1.26 | 361.23 |
| | 100 (99 additions) | Wi-Fi | 7.42 ± 2.64 | 660.94 ± 13.12 | 5.87 ± 1.93 | 674.22 |
| | | Ethernet | 9.19 ± 2.74 | 657.11 ± 18.58 | 7.38 ± 1.91 | 673.67 |
| | 200 (199 additions) | Wi-Fi | 10.04 ± 1.35 | 1316.11 ± 101.74 | 7.56 ± 1.10 | 1333.71 |
| | | Ethernet | 9.63 ± 3.65 | 1290.67 ± 21.73 | 8.39 ± 3.74 | 1308.68 |
| CKKS (addition only) | 50 (49 additions) | Wi-Fi | 15.84 ± 0.50 | 685.10 ± 13.69 | 15.45 ± 2.70 | 716.39 |
| | | Ethernet | 12.84 ± 4.60 | 684.06 ± 18.85 | 12.93 ± 3.84 | 709.83 |
| | 100 (99 additions) | Wi-Fi | 14.51 ± 4.69 | 1406.52 ± 19.63 | 14.47 ± 2.44 | 1435.50 |
| | | Ethernet | 14.77 ± 7.04 | 1338.38 ± 20.42 | 22.30 ± 15.19 | 1375.45 |
| | 200 (199 additions) | Wi-Fi | 16.55 ± 2.30 | 2813.11 ± 122.54 | 28.88 ± 10.43 | 2858.54 |
| | | Ethernet | 13.17 ± 2.74 | 2656.35 ± 22.52 | 18.91 ± 6.00 | 2688.43 |
| CKKS (addition and multiplication) | 50 (49 additions + 1 multiplication) | Wi-Fi | 16.15 ± 5.10 | 799.89 ± 13.60 | 24.44 ± 6.15 | 840.48 |
| | | Ethernet | 12.09 ± 3.31 | 796.48 ± 22.05 | 15.97 ± 0.72 | 824.54 |
| | 100 (99 additions + 1 multiplication) | Wi-Fi | 14.75 ± 1.75 | 1555.08 ± 52.37 | 15.10 ± 0.81 | 1584.93 |
| | | Ethernet | 14.34 ± 1.53 | 1580.64 ± 56.08 | 15.30 ± 2.02 | 1610.28 |
| | 200 (199 additions + 1 multiplication) | Wi-Fi | 20.17 ± 4.87 | 3210.93 ± 327.89 | 20.94 ± 6.15 | 3252.04 |
| | | Ethernet | 14.44 ± 2.87 | 3069.82 ± 14.41 | 16.47 ± 1.93 | 3100.72 |

Note: 6.40 ± 2.04 (mean ± standard deviation)

TABLE VI
COMMUNICATION-LEVEL LATENCY, FRAGMENTATION, AND JITTER CHARACTERISTICS OF HE SCHEMES OVER WI-FI AND ETHERNET

| Scheme | Encrypted Inputs | Medium | Mean RTT (ms) | Jitter (ms) | Mean communication latency (ms) | Total number of fragments | Fragmentation time at both sender and receiver (ms) | Reassembly time at both sender and receiver (ms) |
|---|---|---|---|---|---|---|---|---|
| Baseline (without HE: Message sent: 4 bytes; received: 6 bytes) | 200 (not encrypted) | Wi-Fi | 4.14 | 3.95 | - | 1 | - | - |
| | | Ethernet | 3.81 | 1.12 | - | 1 | - | - |
| BFV (addition only) | 50 | Wi-Fi | 3289.62 | 66.40 | 3088.90 | 95 | 0.07 ± 0.02 | 0.46 ± 0.35 |
| | | Ethernet | 3191.32 | 31.99 | 2997.56 | | 0.06 ± 0.02 | 0.43 ± 0.35 |
| | 100 | Wi-Fi | 3430.66 | 13.10 | 3072.89 | | 0.06 ± 0.02 | 0.44 ± 0.38 |
| | | Ethernet | 3420.37 | 19.57 | 3066.60 | | 0.07 ± 0.02 | 0.49 ± 0.40 |
| | 200 | Wi-Fi | 3769.45 | 43.13 | 3076.99 | | 0.07 ± 0.02 | 0.50 ± 0.37 |
| | | Ethernet | 3767.48 | 25.02 | 3080.75 | | 0.07 ± 0.02 | 0.40 ± 0.27 |
| BGV (addition only) | 50 | Wi-Fi | 9590.36 | 220.96 | 9225.09 | 284 | 0.24 ± 0.09 | 1.20 ± 0.86 |
| | | Ethernet | 9642.65 | 25.19 | 9281.42 | | 0.20 ± 0.07 | 1.15 ± 0.84 |
| | 100 | Wi-Fi | 10003.95 | 262.85 | 9329.74 | | 0.20 ± 0.06 | 1.15 ± 0.93 |
| | | Ethernet | 9924.82 | 164.72 | 9251.15 | | 0.21 ± 0.08 | 1.17 ± 0.82 |
| | 200 | Wi-Fi | 10618.15 | 131.54 | 9284.44 | | 0.22 ± 0.06 | 1.24 ± 0.92 |
| | | Ethernet | 10549.65 | 166.65 | 9240.97 | | 0.22 ± 0.10 | 1.12 ± 0.76 |
| CKKS (addition only) | 50 | Wi-Fi | 19269.65 | 110.50 | 18553.26 | 566 | 0.43 ± 0.14 | 3.03 ± 2.72 |
| | | Ethernet | 19176.56 | 84.87 | 18466.73 | | 0.41 ± 0.12 | 2.73 ± 2.74 |
| | 100 | Wi-Fi | 19985.50 | 55.49 | 18550.00 | | 0.46 ± 0.20 | 3.00 ± 2.72 |
| | | Ethernet | 19764.37 | 130.11 | 18388.91 | | 0.57 ± 0.38 | 2.44 ± 2.21 |
| | 200 | Wi-Fi | 21569.25 | 687.71 | 18710.71 | | 0.50 ± 0.16 | 3.28 ± 2.96 |
| | | Ethernet | 21179.45 | 119.41 | 18491.02 | | 0.43 ± 0.13 | 2.83 ± 2.49 |
| CKKS (addition and multiplication) | 50 | Wi-Fi | 22253.20 | 59.90 | 21412.71 | 754 | 0.57 ± 0.20 | 4.38 ± 5.99 |
| | | Ethernet | 22368.36 | 17.70 | 21543.82 | | 0.40 ± 0.10 | 4.90 ± 6.05 |
| | 100 | Wi-Fi | 23122.92 | 85.62 | 21537.98 | | 0.49 ± 0.21 | 4.75 ± 6.12 |
| | | Ethernet | 23171.26 | 73.18 | 21560.98 | | 0.44 ± 0.09 | 5.78 ± 8.46 |
| | 200 | Wi-Fi | 24972.31 | 477.79 | 21720.27 | | 0.51 ± 0.09 | 4.97 ± 6.92 |
| | | Ethernet | 24630.92 | 32.56 | 21530.20 | | 0.39 ± 0.12 | 4.79 ± 6.47 |

Note: 0.07 ± 0.02 (mean ± standard deviation)

However, one key difference lies in jitter behavior. Wi-Fi demonstrates significantly higher variability in packet delivery timing, with jitter reaching as high as 687.71 ms in the CKKS scenario with 200 encrypted inputs. By contrast, Ethernet consistently maintains jitters below 170 ms across all tested configurations. This disparity in stability has direct implications for packet pacing strategies. In the Wi-Fi setup, packet loss and program halts were observed unless packets were sent with a delay of at least 1.2 times the RTT, leading to inter-packet (inter-trial) spacing as long as 30 seconds for high-load CKKS tests. On the other hand, Ethernet's deterministic environment enabled fixed 100-ms inter-packet delays without any packet loss, supporting continuous data transmission under high volume conditions. From an ITS deployment perspective, these results suggest that Ethernet-based backhaul is well suited for real-time or near-real-time encrypted analytics (e.g., adaptive signal control or coordinated RSU aggregation), while Wi-Fi links are more appropriate for periodic or non-time-critical analytics (e.g., regional speed profiling or eco-driving feedback).

### *C. Latency and Overhead Analysis Across HE Schemes*

As Ethernet exhibited nearly identical communication latency and fragmentation behavior, this subsection reports only the Wi-Fi results, which are representative of both network conditions. Fig. 2 compares the total communication latency (Wi-Fi) across HE schemes. The bar heights correlate strongly with ciphertext size and fragment count. BFV exhibits the lowest latency (~3 s), followed by BGV (~9.2 s), while CKKS (with multiplication) peaks at over 21 s. This reflects the direct relationship between ciphertext size and transmission delay. Notably, these ciphertext sizes are a function of the polynomial ring ($n$) used in each scheme, i.e., 4096 for BFV, 8192 for BGV, and 16384 for CKKS, which significantly impacts serialization size and the resulting number of fragments. This figure reinforces a key insight- communication latency, driven by fragmentation, is the dominant bottleneck in post-quantum HE-based I2I systems. Even relatively small ciphertexts like BFV (132 KB) generate

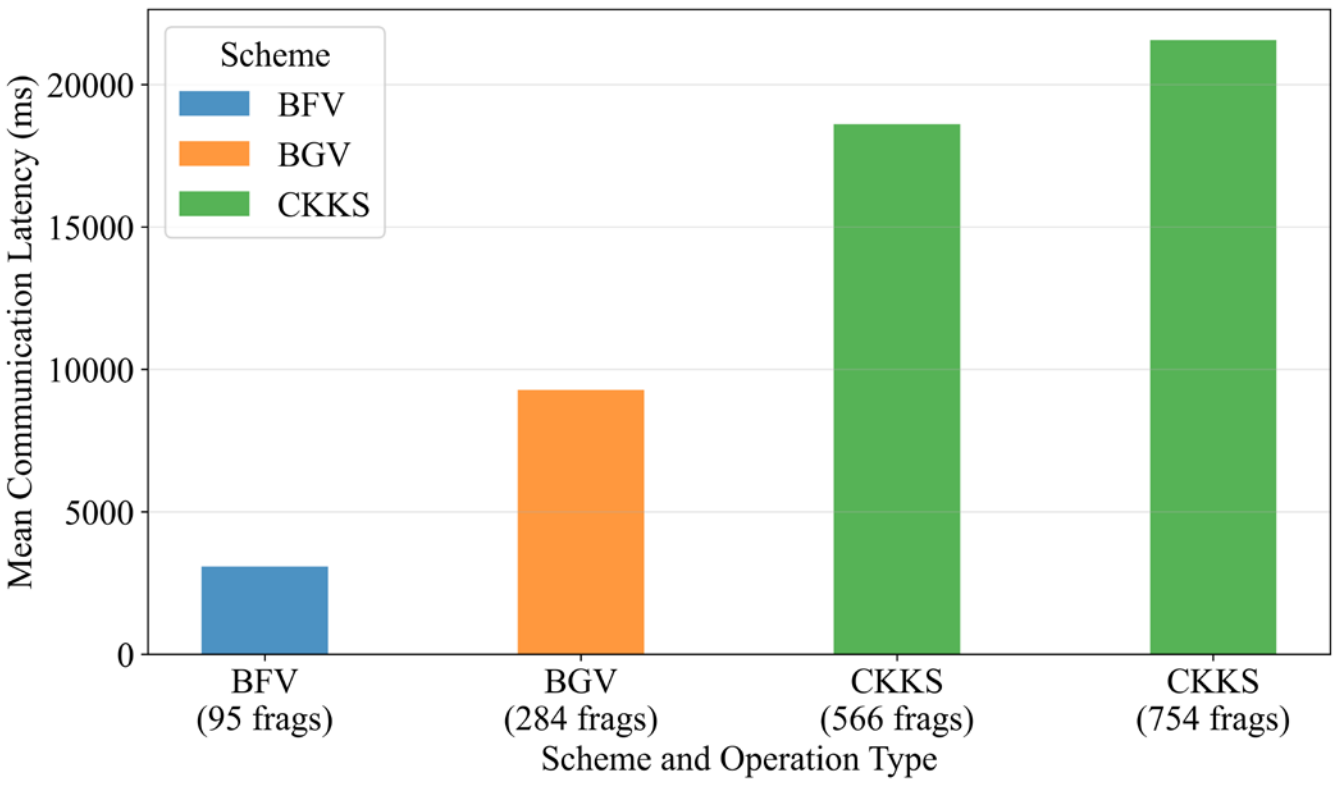

Fig. 2. Mean communication latency by HE schemes and parameters (Wi-Fi).

approximately 100 fragments, stressing Wi-Fi under tight pacing intervals.

Fig. 3 presents the mean HE computation times at the receiver across different workloads in Wi-Fi scenario. As the number of additions increases, computation time gradually rises for all schemes. When a single multiplication is added to the CKKS workload, the total processing time increases further, e.g., from approximately 2.8 s to 3.2 s for 200 encrypted inputs, highlighting the additional computational cost associated with multiplicative operations in CKKS.

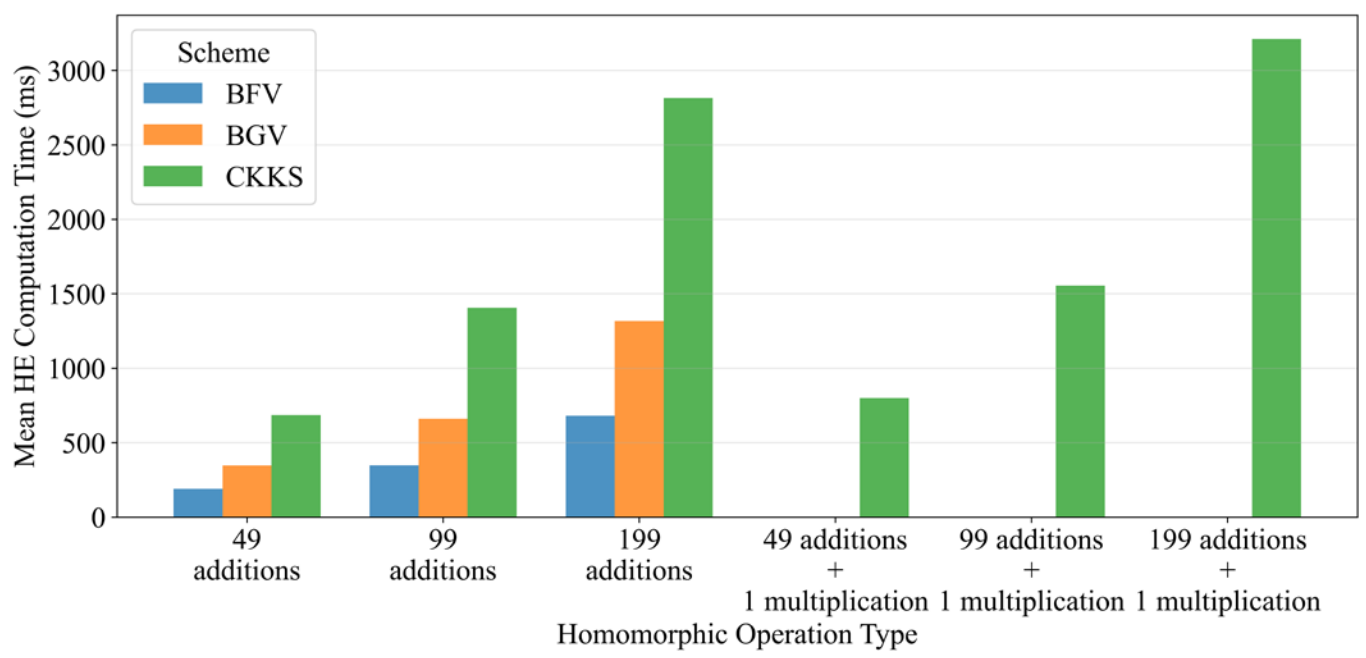

Fig. 3. Mean HE computation time by number of additions and multiplicative depth (Wi-Fi).

### *D. Feasibility and Suitability of Post-Quantum HE Schemes for ITS Backhaul Applications*

The combined experimental results suggest that HE can support a range of privacy-preserving ITS backhaul applications, provided that the latency requirements of the use case align with the observed system performance. For congestion monitoring and intersection control (e.g., TM03), the BFV scheme, with a total latency of approximately 3 seconds, can accommodate sub-5-second encrypted updates. Its low encryption and decryption overhead minimize computational load on RSUs, making it suitable for localized data aggregation and real-time intersection analytics. The BGV scheme, with end-to-end latency around 9 seconds, supports regional aggregation tasks, such as corridor-level flow estimation or traffic-density heatmap generation (e.g., TM07). This performance allows periodic updates on 10-30 second cycles while maintaining strong privacy guarantees.

In contrast, the CKKS scheme, with 21-32 seconds of total latency depending on input size and arithmetic depth, is less suited to near-real-time operations. However, its ability to process encrypted floating-point values makes it practical for longer-interval analytics, such as minute-scale eco-driving feedback, historical average-speed computation, or fuel-efficiency monitoring. While parameter tuning could reduce CKKS latency by lowering polynomial ring sizes, such changes would compromise the 128-bit post-quantum security target.

However, it is important to note that none of the tested HE schemes, under current parameterizations of at least 128-bit security and network constraints, meet the stringent end-to-end latency requirements of real-time, safety-critical vehicle-level applications, such as forward collision warning or emergency braking systems, which typically demand round-trip latencies below 100 ms [55]. The experimental latency values observed, ranging from 3 to over 30 seconds depending on scheme and workload, fall well outside this threshold. As such, the direct use of HE in latency-critical vehicular safety systems remains impractical without significant advances in ciphertext compression, high-throughput networking, multi-threaded data transmission, and specialized hardware acceleration.

## V. CONCLUSION AND FUTURE WORK

This study presents the first real-world experimental benchmarking of post-quantum secure HE schemes (BFV, BGV, and CKKS) for privacy-preserving I2I data aggregation in ITS. Unlike prior work limited to simulation or theoretical modeling, this research implemented and evaluated HE-based encrypted analytics between RSU and Cloud/Data Center entities using real hardware and both Wi-Fi and Ethernet communication. Two representative ITS workloads were tested: encrypted summation (e.g., congestion detection and queue estimation) and encrypted averaging (e.g., speed harmonization and eco-driving analytics). All schemes were implemented using the OpenFHE library and executed on real hardware with application-level measurements of end-to-end computational and communication delay.

Results demonstrate that even with conservative parameters ensuring 128-bit post-quantum security, HE is practically viable for latency-tolerant ITS backhaul applications. The BFV scheme supports sub-5-second encrypted updates suitable for intersection-level analytics such as TM03 (Traffic Signal Control). BGV, with moderate latency, enables regional aggregation applications, such as TM07 (Regional Traffic Management) or TM02 (Probe-based Surveillance). Although CKKS introduces higher latency, it remains suitable for minute-level encrypted floating-point analytics like TM21 (Speed Harmonization) or long-term eco-driving evaluation. However, none of the tested schemes currently meet the sub-second response time required for safety-critical ITS functions, reaffirming that HE's immediate value lies in privacy-preserving infrastructure analytics rather than direct vehicle control loops.

Future work will focus on reducing end-to-end delay across the HE-enabled RSU-Cloud communication pipeline through optimization at the algorithmic, network, and implementation levels. At the algorithmic level, lightweight parameter tuning and hybrid encryption models will be explored to balance security and efficiency. Network-level efforts will target adaptive packet pacing, optimized fragmentation, and protocol tuning to minimize latency and jitter. Implementation-level improvements will leverage hardware acceleration using GPUs, FPGAs, or embedded co-processors to parallelize encryption and computation. Field trials with commercial RSUs and cloud computing platforms featuring LTE/5G/Ethernet backhaul will validate performance under real-world conditions, while integration with secure multi-party computation and differential privacy will further enhance collaborative and privacy-preserving ITS analytics.

## ACKNOWLEDGMENT

This work is based upon the work supported by the National Center for Transportation Cybersecurity and Resiliency (TraCR) (a U.S. Department of Transportation National University Transportation Center) headquartered at Clemson University, Clemson, South Carolina, USA. Any opinions, findings, conclusions, and recommendations expressed in this material are those of the author(s) and do not necessarily reflect the views of TraCR, and the U.S. Government assumes no liability for the contents or use thereof.

## REFERENCES

[1] T. Yoshizawa *et al.*, “A Survey of Security and Privacy Issues in V2X Communication Systems,” *ACM Comput. Surv.*, vol. 55, no. 9, pp. 1–36, Sept. 2023, doi: 10.1145/3558052.

[2] A. Alnasser, H. Sun, and J. Jiang, “Cyber security challenges and solutions for V2X communications: A survey,” *Computer Networks*, vol. 151, pp. 52–67, Mar. 2019, doi: 10.1016/j.comnet.2018.12.018.

[3] U.S. Department of Transportation ITS Standards Program, “SAE J2735 – Dedicated Short Range Communications (DSRC) Message Set Dictionary.” [Online]. Available: https://www.standards.its.dot.gov/Factsheets/Factsheet/71

[4] “SCMS Fact Sheet,” University of Virginia, Connected Vehicle Pooled Fund Study (CVPFS), Mar. 2022. [Online]. Available: https://www.engineering.virginia.edu/sites/default/files/Connected-Vehicle-PFS/Resources/CV%20PFS%20SCMS%20Fact%20Sheet%2003072022.pdf

[5] S. Bitam and A. Mellouk, “ITS-cloud: Cloud computing for Intelligent transportation system,” in *2012 IEEE Global Communications Conference (GLOBECOM)*, Anaheim, CA, USA: IEEE, Dec. 2012, pp. 2054–2059. doi: 10.1109/GLOCOM.2012.6503418.

[6] M. S. Salek *et al.*, “A Review on Cybersecurity of Cloud Computing for Supporting Connected Vehicle Applications,” *IEEE Internet Things J.*, vol. 9, no. 11, pp. 8250–8268, June 2022, doi: 10.1109/JIOT.2022.3152477.

[7] National Highway Traffic Safety Administration, “Privacy Impact Assessment for the Vehicle-to-Vehicle (V2V) Notice of Proposed Rulemaking (NPRM),” U.S. Department of Transportation, Washington, DC, Dec. 2016. [Online]. Available: https://www.transportation.gov/sites/dot.gov/files/docs/Privacy%20-%20NHTSA%20-%20V2V%20NPRM%20-%20PIA%20-%20Approved%20-%20122016.pdf

[8] J. Petit, F. Schaub, M. Feiri, and F. Kargl, “Pseudonym Schemes in Vehicular Networks: A Survey,” *IEEE Commun. Surv. Tutorials*, vol. 17, no. 1, pp. 228–255, 2015, doi: 10.1109/COMST.2014.2345420.

[9] National Institute of Standards and Technology, “Advanced encryption standard (AES),” National Institute of Standards and Technology, Gaithersburg, MD, NIST FIPS 197, Nov. 2001. doi: 10.6028/NIST.FIPS.197.

[10] National Institute of Standards and Technology (US), “Module-lattice-based key-encapsulation mechanism standard,” National Institute of Standards and Technology (U.S.), Washington, D.C., NIST FIPS 203, Aug. 2024. doi: 10.6028/NIST.FIPS.203.

[11] C. Marcolla, V. Sucasas, M. Manzano, R. Bassoli, F. H. P. Fitzek, and N. Aaraj, “Survey on Fully Homomorphic Encryption, Theory, and Applications,” *Proc. IEEE*, vol. 110, no. 10, pp. 1572–1609, Oct. 2022, doi: 10.1109/JPROC.2022.3205665.

[12] V2X Core Technical Committee, *On-Board System Requirements for V2V Safety Communications*. doi: 10.4271/J2945/1_202004.

[13] G. Wang *et al.*, “Towards Efficient Privacy-Preserving Keyword Search for Outsourced Data in Intelligent Transportation Systems,” 2025, *Elsevier BV*. doi: 10.2139/ssrn.5180330.

[14] D. Palma, P. L. Montessoro, M. Loghi, and D. Casagrande, “A Privacy-Preserving System for Confidential Carpooling Services Using Homomorphic Encryption,” *Advanced Intelligent Systems*, vol. 7, no. 5, p. 2400507, May 2025, doi: 10.1002/aisy.202400507.

[15] B. Mi, J. Zhou, D. Huang, and Y. Weng, “Privacy-Preserving Data Processing Method for IoV Based on Homomorphic Conjugacy Search Problem,” *IEEE Trans. Intell. Transport. Syst.*, vol. 25, no. 7, pp. 7374–7387, July 2024, doi: 10.1109/tits.2024.3351837.

[16] H. Karim and D. B. Rawat, “TollsOnly Please—Homomorphic Encryption for Toll Transponder Privacy in Internet of Vehicles,” *IEEE Internet Things J.*, vol. 9, no. 4, pp. 2627–2636, Feb. 2022, doi: 10.1109/jiot.2021.3056240.

[17] Y. Ameur and S. Bouzefrane, “Enhancing privacy in VANETs through homomorphic encryption in machine learning applications,” *Procedia Computer Science*, vol. 238, pp. 151–158, 2024, doi: 10.1016/j.procs.2024.06.010.

[18] J. Fan and F. Vercauteren, “Somewhat Practical Fully Homomorphic Encryption,” 2012, [Online]. Available: https://eprint.iacr.org/2012/144

[19] Z. Brakerski, C. Gentry, and V. Vaikuntanathan, “(Leveled) Fully Homomorphic Encryption without Bootstrapping,” *ACM Trans. Comput. Theory*, vol. 6, no. 3, pp. 1–36, July 2014, doi: 10.1145/2633600.

[20] Z. Brakerski, C. Gentry, and V. Vaikuntanathan, “Fully Homomorphic Encryption without Bootstrapping.” 2011. [Online]. Available: https://eprint.iacr.org/2011/277

[21] J. H. Cheon, A. Kim, M. Kim, and Y. Song, “Homomorphic Encryption for Arithmetic of Approximate Numbers,” 2016, [Online]. Available: https://eprint.iacr.org/2016/421

[22] V. Lyubashevsky, C. Peikert, and O. Regev, “On Ideal Lattices and Learning with Errors over Rings,” *J. ACM*, vol. 60, no. 6, pp. 1–35, Nov. 2013, doi: 10.1145/2535925.

[23] P. Paillier, “Public-Key Cryptosystems Based on Composite Degree Residuosity Classes,” in *Advances in Cryptology — EUROCRYPT ’99*, vol. 1592, J. Stern, Ed., in Lecture Notes in Computer Science, vol. 1592. , Berlin, Heidelberg: Springer Berlin Heidelberg, 1999, pp. 223–238. doi: 10.1007/3-540-48910-X_16.

[24] S. O. Ogundoyin, “An anonymous and privacy-preserving scheme for efficient traffic movement analysis in intelligent transportation system,” *Security and Privacy*, vol. 1, no. 6, Nov. 2018, doi: 10.1002/spy2.50.

[25] F. Farokhi, I. Shames, and K. H. Johansson, “Private routing and ride-sharing using homomorphic encryption,” *IET Cyber-Phy Sys Theory & Ap*, vol. 5, no. 4, pp. 311–320, Dec. 2020, doi: 10.1049/iet-cps.2019.0042.

[26] W. Ren *et al.*, “Privacy-preserving using homomorphic encryption in Mobile IoT systems,” *Computer Communications*, vol. 165, pp. 105–111, Jan. 2021, doi: 10.1016/j.comcom.2020.10.022.

[27] A. Sultan *et al.*, “A Novel Image-Based Homomorphic Approach for Preserving the Privacy of Autonomous Vehicles Connected to the Cloud,” *IEEE Trans. Intell. Transport. Syst.*, pp. 1–13, 2022, doi: 10.1109/TITS.2022.3219591.

[28] Y. Liu, X. Xiao, F. Kong, H. Zhang, and J. Yu, “Towards efficient privacy-preserving conjunctive keywords search over encrypted cloud data,” *Future Generation Computer Systems*, vol. 166, p. 107716, May 2025, doi: 10.1016/j.future.2025.107716.

[29] A. Lakhan, T.-M. Groenli, H. Wu, M. Younas, and G. Ghinea, “A Novel Homomorphic Blockchain Scheme for Intelligent Transport Services in Fog/Cloud and IoT Networks,” *IEEE Trans. Intell. Transport. Syst.*, vol. 26, no. 2, pp. 1914–1929, Feb. 2025, doi: 10.1109/TITS.2024.3493452.

[30] P. W. Shor, “Algorithms for quantum computation: discrete logarithms and factoring,” in *Proceedings 35th Annual Symposium on Foundations of Computer Science*, Santa Fe, NM, USA: IEEE Comput. Soc. Press, 1994, pp. 124–134. doi: 10.1109/SFCS.1994.365700.

[31] S. Gao and K. Yates, “Leveled Homomorphic Encryption Schemes for Homomorphic Encryption Standard.” 2024. [Online]. Available: https://eprint.iacr.org/2024/991

[32] S. Gao, “Efficient Fully Homomorphic Encryption Scheme.” 2018. [Online]. Available: https://eprint.iacr.org/2018/637

[33] B. M. Case, S. Gao, G. Hu, and Q. Xu, “Fully Homomorphic Encryption with k-bit Arithmetic Operations.” 2019. [Online]. Available: https://eprint.iacr.org/2019/521

[34] *openfheorg/openfhe-development*. (July 31, 2025). C++. OpenFHE. Accessed: Aug. 01, 2025. [Online]. Available: https://github.com/openfheorg/openfhe-development

[35] J.-P. Bossuat *et al.*, “Security Guidelines for Implementing Homomorphic Encryption.” 2024. [Online]. Available: https://eprint.iacr.org/2024/463

[36] M. Creeger, "The Rise of Fully Homomorphic Encryption: Often called the Holy Grail of cryptography, commercial FHE is near.," *Queue*, vol. 20, no. 4, pp. 39–60, Aug. 2022, doi: 10.1145/3561800.
[37] A. Hannemann and E. Buchmann, "Is Homomorphic Encryption Feasible for Smart Mobility?," presented at the 18th Conference on Computer Science and Intelligence Systems, Sept. 2023, pp. 523–532. doi: 10.15439/2023F695.
[38] K. Akkaya, V. Baboolal, N. Saputro, S. Uluagac, and H. Menouar, "Privacy-Preserving Control of Video Transmissions for Drone-based Intelligent Transportation Systems," in *2019 IEEE Conference on Communications and Network Security (CNS)*, Washington DC, DC, USA: IEEE, June 2019, pp. 1–7. doi: 10.1109/cns.2019.8802665.
[39] A. Boudguiga, O. Stan, A. Fazzat, H. Labiod, and P.-E. Clet, "Privacy Preserving Services for Intelligent Transportation Systems with Homomorphic Encryption:," in *Proceedings of the 7th International Conference on Information Systems Security and Privacy*, Online Streaming, --- Select a Country ---: SCITEPRESS - Science and Technology Publications, 2021, pp. 684–693. doi: 10.5220/0010349706840693.
[40] G. Costantino, M. De Vincenzi, F. Martinelli, and I. Matteucci, "A Privacy-Preserving Solution for Intelligent Transportation Systems: Private Driver DNA," *IEEE Trans. Intell. Transport. Syst.*, vol. 24, no. 1, pp. 258–273, Jan. 2023, doi: 10.1109/TITS.2022.3217358.
[41] J. Chen, K. Li, and P. S. Yu, "Privacy-Preserving Deep Learning Model for Decentralized VANETs Using Fully Homomorphic Encryption and Blockchain," *IEEE Trans. Intell. Transport. Syst.*, vol. 23, no. 8, pp. 11633–11642, Aug. 2022, doi: 10.1109/tits.2021.3105682.
[42] X. Sun, F. R. Yu, P. Zhang, W. Xie, and X. Peng, "A Survey on Secure Computation Based on Homomorphic Encryption in Vehicular Ad Hoc Networks," *Sensors*, vol. 20, no. 15, p. 4253, July 2020, doi: 10.3390/s20154253.
[43] I. Chillotti, N. Gama, M. Georgieva, and M. Izabachène, "TFHE: Fast Fully Homomorphic Encryption over the Torus." 2018. [Online]. Available: https://eprint.iacr.org/2018/421
[44] {Mazin Abed} Mohammed *et al.*, "Homomorphic federated learning schemes enabled pedestrian and vehicle detection system," *Internet of Things (Netherlands)*, vol. 23, Oct. 2023, doi: 10.1016/j.iot.2023.100903.
[45] Z. Ying, S. Cao, X. Liu, Z. Ma, J. Ma, and R. H. Deng, "PrivacySignal: Privacy-Preserving Traffic Signal Control for Intelligent Transportation System," *IEEE Trans. Intell. Transport. Syst.*, vol. 23, no. 9, pp. 16290–16303, Sept. 2022, doi: 10.1109/TITS.2022.3149600.
[46] K. Wang, C.-M. Chen, M. Shojafar, Z. Tie, M. Alazab, and S. Kumari, "AFFIRM: Provably Forward Privacy for Searchable Encryption in Cooperative Intelligent Transportation System," *IEEE Trans. Intell. Transport. Syst.*, vol. 23, no. 11, pp. 22607–22618, Nov. 2022, doi: 10.1109/TITS.2022.3177899.
[47] United States Department of Transportation (USDOT), "ARC-IT v9.2: Architecture Reference for Cooperative and Intelligent Transportation," ARC-IT. Accessed: June 15, 2025. [Online]. Available: https://www.arc-it.net/
[48] Architecture Reference for Cooperative and Intelligent Transportation, "TM03: Traffic Signal Control." Accessed: Oct. 28, 2025. [Online]. Available: https://www.arc-it.net/html/servicepackages/sp122.html#tab-3
[49] Architecture Reference for Cooperative and Intelligent Transportation, "TM17: Speed Warning and Enforcement." Accessed: Oct. 28, 2025. [Online]. Available: https://www.arc-it.net/html/servicepackages/sp135.html#tab-3
[50] Architecture Reference for Cooperative and Intelligent Transportation, "TM21: Speed Harmonization." Accessed: Oct. 28, 2025. [Online]. Available: https://www.arc-it.net/html/servicepackages/sp68.html#tab-3
[51] Architecture Reference for Cooperative and Intelligent Transportation, "TM02: Vehicle-Based Traffic Surveillance." Accessed: Oct. 28, 2025. [Online]. Available: https://www.arc-it.net/html/servicepackages/sp87.html#tab-3
[52] Architecture Reference for Cooperative and Intelligent Transportation, "MC06: Work Zone Management." Accessed: Oct. 28, 2025. [Online]. Available: https://www.arc-it.net/html/servicepackages/sp148.html#tab-3
[53] Architecture Reference for Cooperative and Intelligent Transportation, "TM07: Regional Traffic Management." Accessed: Oct. 28, 2025. [Online]. Available: https://www.arc-it.net/html/servicepackages/sp126.html#tab-3
[54] "MK6," Cohda Wireless. Accessed: Oct. 28, 2025. [Online]. Available: https://cohdawireless.com/solutions/mk6/
[55] G. Karagiannis *et al.*, "Vehicular Networking: A Survey and Tutorial on Requirements, Architectures, Challenges, Standards and Solutions," *IEEE Commun. Surv. Tutorials*, vol. 13, no. 4, pp. 584–616, 2011, doi: 10.1109/SURV.2011.061411.00019.

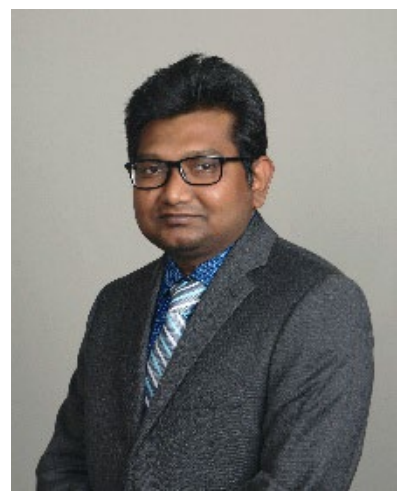

ABDULLAH AL MAMUN (Student Member, IEEE) received the M.S. and Ph.D. degrees in Environmental Engineering and Earth Sciences from Clemson University, Clemson, SC, USA, in 2020 and 2022, respectively.

He is currently pursuing his second Ph.D. in Civil Engineering at Clemson University, Clemson, SC, USA. His research interests include cybersecurity and resiliency of Intelligent Transportation Systems (ITS) and post-quantum cryptography.

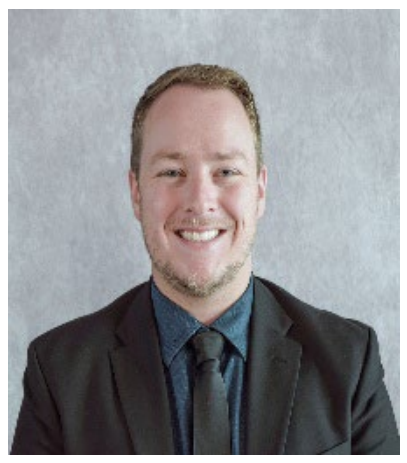

KYLE YATES received the B.S. degree in applied mathematics from San Diego State University, San Diego, California, USA, in 2020, and the M.S. degree in mathematical sciences from Clemson University, Clemson, SC, in 2022, where he is currently pursuing the Ph.D. degree. His primary research interests include the mathematics of post-quantum cryptography and homomorphic encryption.

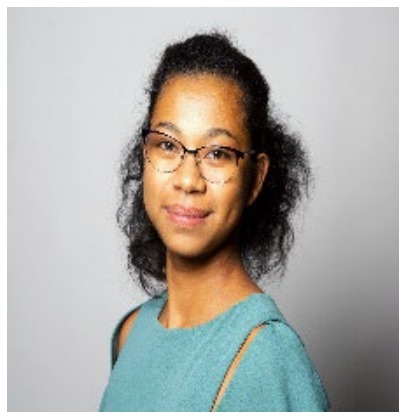

ANTSA PIERROTTET is a Ph.D. student in mathematics at Clemson University, focusing on post-quantum cryptography. She earned her undergraduate degree in mathematics education from the University of Antananarivo, Madagascar, and holds two Master of Science degrees in mathematics from AIMS, South Africa, and Clemson University, South Carolina, USA.

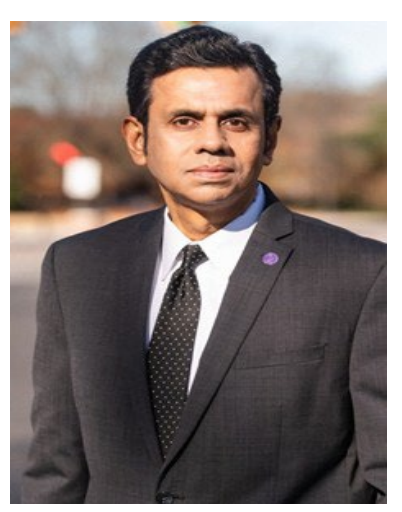

MASHRUR CHOWDHURY (Senior Member, IEEE) received the Ph.D. degree in civil engineering from the University of Virginia, in 1995.

Currently, he is the Eugene Douglas Mays Chair of transportation with the Glenn Department of Civil Engineering, Clemson University. He also serves as the Founding Director of the USDOT-sponsored National Center for Transportation Cybersecurity and Resiliency (TraCR). His current research interests include evolving realms of sensing, communications, computing, cybersecurity, and cyber-resiliency, all with the goal of establishing a secure and resilient IoT environment for smart cities and regions.

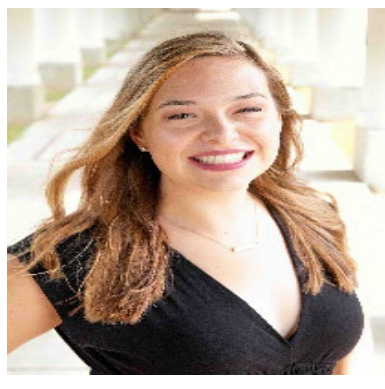

RYANN CARTOR received her Ph.D. in applied mathematics from the University of Louisville in 2019. She is currently an Assistant Professor in the School of Mathematical and Statistical Sciences at Clemson University. Her research interests include Post-Quantum Cryptography, with an emphasis in Multivariate and Code-Based Cryptography.

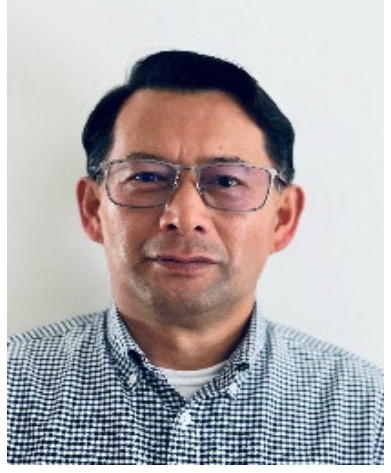

SHUHONG GAO completed his BS (1983) and MS (1986) degrees in Mathematics at Sichuan University, China, and earned his PhD in Combinatorics and Optimization from the University of Waterloo, Canada, in 1993.

Currently, he is a Professor in the School of Mathematical and Statistical Sciences at Clemson University. Dr. Gao has published over 80 papers in areas including combinatorial design theory, finite fields, coding theory, cryptography, symbolic computation, and computational algebraic geometry. He has supervised 17 PhD students, and his research has been supported by grants from the NSA, NSF, and ONR.